\documentclass{article}
\usepackage[utf8]{inputenc}

\title{The virial theorem for non-differentiable dynamical paths in resolution-scale relativity}
\author{Tugdual LeBohec}
\date{September $13^{th}$ 2022}
\begin{document}

\maketitle
\abstract{The virial theorem is established in the framework of resolution-scale relativity for stochastic dynamics characterized by a diffusion constant $\mathcal D$. It only relies on a simple time average just like the classical virial theorem, while the quantum mechanical virial theorem involves the expectation values of the observables. Nevertheless, by the emergence of a {\it quantum-like potential} term, the resolution-scale relativity virial theorem also encompasses quantum mechanical dynamics under the identification $\hbar \leftrightarrow 2m\mathcal D$ . This provides an illustration of the scale relativistic approach to the foundation of standard quantum mechanics. Furthermore, it is pointed out that, if the resolution-scale relativity principle is implemented in macroscopic systems that are complex and/or chaotic, then the application of the classical virial theorem in the analysis of the dynamics of astrophysical systems neglects the contribution from a resolution-scale relativistic quantum-like potential. It is shown that this quantum-like potential could account for some fraction of the dark matter hypothesis.}

\section{Introduction}

The Ehrenfest theorem \cite{ehrenfest1927} reveals the precise meshing of quantum mechanics with classical mechanics. As such, it makes the wave function appear as a {\it container} of the specifically quantum dynamics, which it separates from the classical dynamics.  The latter is {\it external} to the wave function, of which only the expectation values are then relevant. 

Some systems or problems are traditionally treated as fully classical, others as fully quantum, and yet others as a combination of classical and quantum dynamics. Examples of the combined case come from the Born-Oppenheimer approximation \cite{bornoppenheimer1927}, in which the system's wave-function is factorized according to the different time scales at play. Sometimes, the component with the slowest time evolution is treated classically \cite{frank2020}. This illustrates the certain freedom we have in the choice of what goes {\it inside} the quantum wave-function {\it container} and what remains classical on the {\it outside}.  The choice is generally driven by the practicality of the model and the level of precision required for the predictions it provides. However, the most fundamental description of a system is understood to be fully quantum mechanical.  

This connection between the seemingly separated classical and quantum domains is clearly illustrated by the virial theorem in dissipation free systems. In classical mechanics, the virial theorem \cite{clausius1870} applies to bound systems, and equates the time averaged kinetic energy to the time average of a function of the interactions between the system constituents. In 1930, Vladimir Fock \cite{fock1930} noticed that the Ehrenfest theorem implies the same equality in quantum mechanics with the classical time averaging replaced with the time averaging of the  quantum expectation values of the corresponding quantities. This is not only saying that the classical virial theorem can be obtained from quantum mechanics by taking a few expectation values. It is also, and more interestingly, saying that the quantum dynamics {\it internal} to the wave function complies with an extended version of the virial theorem, in which the taking of expectation values plays the role of a quantum extension to the classical time averaging. In this article, I wish to investigate this connection in the context of resolution-scale relativity \cite{nottale1993, nottale2011}.

The resolution-scale relativistic formulation of standard quantum mechanics interprets wave-functions as time cross-sections of bundles of innumerable and non-differentiable dynamical paths \cite{nottale2007, MHTeh2017, saeed2020}.  In this approach, the quantum behavior emerges from the non-differentiable nature of the dynamical paths. In particular, there is no equivalent to wave functions as {\it containers} separating the quantum from the classical dynamics. Instead, the transition between domains with predominating quantum and classical dynamics respectively occurs by changing the resolution-scale with which the paths bundles are considered. As the dynamical paths are innumerable and indistinguishable, the system cannot be regarded as following any specific one of them. When the resolution-scale  is degraded to such a coarse level that the dynamical paths lose their non-differentiable nature, the path bundles collapse to the actual trajectories of classical mechanics. The differentiable smoothed out trajectories account for the classical behavior while the non-differentiable residuals accounts for the quantum dynamics. The two coexist, each playing a preponderant role in its respective observation resolution-scale domain. Both are described together by a single equation of dynamics. The resolution-scale of transition between the differentiable and predominantly classical regime and the non-differentiable and predominantly quantum regime  corresponds to the de\,Broglie wave-length. The connection made by the virial theorem between the specifically quantum and classical dynamics and the non-differentiable paths based formulation of quantum mechanics makes it interesting to review the virial theorem from the resolution-scale relativity approach, which is precisely the goal of this article.  
 
In order to do this, Section \ref{virialqc} is used to briefly review the classical and quantum virial theorems. This article does not intend to review in details the non-differential paths dynamics and the foundation of quantum mechanics it provides in the resolution-scale relativity framework. For this, the reader may consult publications by L.Nottale \cite{nottale1993, nottale2011} or by the author of the present article \cite{MHTeh2017, saeed2020}. However, in order to set the notations used in the rest of the article, Section \ref{srdynamics} provides a very brief overview of non-differentiable dynamics. Then, Section \ref{srvirialsec} proceeds with establishing the virial theorem from a resolution-scale relativity approach. Finally, Section \ref{sumconc} provides a summary and conclusions clarifying the nature of the resolution-scale relativistic virial theorem and envisions possible implications if the scale relativity principle is implemented in the dynamics of complex and/or chaotic systems.

\section{Classical and quantum virial}\label{virialqc}
In classical mechanics, for a closed and isolated system of $N$ mutually interacting particles with respective masses $m_k$, positions ${\bf r}_k$, and momenta ${\bf p}_k$, the virial is a quantity homogeneous to an action and defined as $G=\sum\limits_{k=1}^N{\bf r}_k\cdot {\bf p}_k$. The virial theorem then arises from considering the time average of the virial energy, which is the time derivative of the virial: ${{dG}\over{dt}}=2T_{Tot}+\sum\limits_{k=1}^N {\bf r}_k \cdot {\bf F}_k$, where $T_{Tot}=\sum\limits_{k=1}^N\frac{{\bf p}_k^2}{2m_k}$ is the kinetic energy of the whole system, and ${\bf F}_k$ is the force acting on the $k^{th}$ particle.  In the case of a finite and bound system of particles globally at rest, both ${\bf r}_k$ and  ${\bf p}_k$ are bound and so is the virial. Hence, in the limit of large times, the time average of the virial energy vanishes or $\overline{\left( {{dG}\over{dt}}\right)}=0$, where $\overline{\left(\cdots\right)}$ denotes the time average. And so, the classical mechanics virial theorem takes the form:
\begin{equation}\label{classvirial}
2\overline{\left( T_{Tot}\right)}=-\overline{\left( \sum\limits_{k=1}^N {\bf r}_k \cdot {\bf F}_k\right)}.
\end{equation}
When the force ${\bf F}_k$ results from a pair-wise central force interaction between the particles with a magnitude following a power law of the distance, the virial theorem provides a relation between the system's time averaged kinetic and potential energies. 
This is of great use in the study of gravitationally bound astrophysical systems as the measurement of the systems kinematics can be used to probe the interaction between the system's constituents or the possible presence of unseen material contributing to the gravitational potential \cite{zwicky1937}. 

From here, going toward quantum mechanics, we can redefine  the classical virial by replacing positions and momenta  with their respective quantum expectation values $G_C=\sum\limits_{k=1}^N\langle{\bf r}_k\rangle\cdot\langle {\bf p}_k\rangle$. With the Hamiltonian $H=\sum\limits_{k=1}^N \frac{{\bf p}_k^2}{2m_k}+ V_{Tot}$ (where $\hat V_{Tot}$ is a potential, from which the forces ${\bf F}_k$ derive in position representation), the Herenfest theorem implies that $\langle{\bf r}_k\rangle$ and $\langle {\bf p}_k\rangle$ evolve in time following the prescription of classical mechanics and so we obtain a new form of the virial theorem as (See Appendix A):  
\begin{equation}\label{expectvirial}
\overline{\left( \sum\limits_{k=1}^N \frac{\langle{\bf p}_k\rangle^2}{m_k}\right)}=-\overline{\left( \sum\limits_{k=1}^N \langle{\bf r}_k\rangle \cdot \langle {\bf F}_k\rangle\right)}.
\end{equation}
Because of the role played byt the Ehrenfest theorem, the virial theorems expressed by Equations \ref{classvirial} and \ref{expectvirial} are in fact the same. It should however be noted that the left-hand side of Equation \ref{expectvirial} is not the time average of the expectation value of the kinetic energy and the right hand side is not either the time average of the expectation value of $\sum\limits_{k=1}^N {\bf r}_k \cdot {\bf F}_k$. So one does not go from the classical virial theorem expressed by Equation \ref{classvirial} to the quantum based expression of the same theorem expressed by Equation \ref{expectvirial} by simply expanding the time averaging to include taking the expectation values in both sides of Equation \ref{classvirial}. 

Instead, we may now consider an actually quantum virial $G_Q=\frac{1}{2}\sum\limits_{k=1}^N\langle{\bf r}_k\cdot {\bf p}_k+{\bf p}_k\cdot {\bf r}_k\rangle$. Using Ehrenfest theorem again (See Appendix B), the quantum virial theorem is obtained as:
\begin{equation}2\overline{\left(\langle \hat T_{Tot}\rangle\right)}=-\overline{\left( \langle\sum\limits_{k=1}^N {\bf r}_k \cdot {\bf F}_k\rangle\right)},
\label{quantvirialeq}
\end{equation}
where we use $\hat T_{Tot}$ to represent the total quantum kinetic energy. This was originally noticed by Vladimir Fock \cite{fock1930} in 1930. The quantum virial theorem expressed by Equation \ref{quantvirialeq} has the same form as the classical virial theorem of Equation \ref{classvirial} provided the time averaging $\overline{\left(\cdots\right)}$ in the latter is extended to first include taking the quantum expectation values $\langle \cdots\rangle$ of the respective observables, which includes the contribution of dynamics {\it internal} to the wave function. In fact, when considering the system to be in a quantum stationary state, the time averaging becomes superfluous and we obtain a relation between expectation values $2\langle \hat T_{Tot}\rangle = \langle\sum\limits_{k=1}^N {\bf r}_k \cdot {\bf F}_k\rangle$, which, when thought of as an averaging over an {\it internal ergodic dynamics}, is equivalent to the classical form of the virial theorem: $2\overline{\left(T_{Tot}\right)} = \overline{\left(\sum\limits_{k=1}^N {\bf r}_k \cdot {\bf F}_k\right)}$. At the same time, and in complementarity, in stationary states, the classical virial theorem for $G_C$ (Equation \ref{expectvirial}) becomes degenerate as both sides evaluate to zero since stationarity implies both $\langle{\bf p}_k\rangle=0$ and $\langle{\bf F}_k\rangle=0$, while it is clear that the expectation value of the kinetic and potential energies are non-zero.  This clearly and specifically illustrates the partitioning between classical and quantum dynamics.

\section{Non-differentiable paths dynamics}\label{srdynamics}
The resolution-scale relativity principle introduced by L.\,Nottale \cite{nottale1993} amounts to an extension of the relativity principle to resolution-scale transformations.  Here, I do not intend to provide a coverage of the principle based foundation of standard quantum mechanics provided by Nottale's resolution-scale relativity. Instead, I briefly outline the main steps just to introduce notations while refraining from commenting on the interpretation \cite{nottale1993, nottale2007,nottale2011,MHTeh2017, saeed2020}. 

The enforcement of the resolution-scale relativity principle in the foundation of a theory of dynamics leads to include non-differentiable dynamical paths in the analysis. This requires the use of finite time steps instead of the infinitesimal ones of standard calculus and it also requires to abandon the symmetry between forward and backward microscopic time steps. Hence, we define
\begin{eqnarray*}
\frac{d_+}{dt}\bigg |_{\delta t}{\bf r}&=\frac{{\bf r}(t+\delta t,\delta t)-{\bf r}(t,\delta t)}{\delta t}&={\bf v}_+(t,\delta t){~~~~\delta t>0},\\
\frac{d_-}{dt}\bigg |_{\delta t}{\bf r}&=\frac{{\bf r}(t,\delta t)-{\bf r}(t-\delta t,\delta t)}{\delta t}&={\bf v}_-(t,\delta t){~~~~\delta t<0}.
\end{eqnarray*}

So the velocity becomes both double valued and resolution-scale dependent with $\delta t$ setting the resolution-scale in use. In order to efficiently deal with the double-valuedness, without any loss of generality, it is convenient to take advantage of complex numbers by defining a single {\it complex time-differential operator} \cite{nottale2011}, 
\begin{eqnarray*}
\frac{\hat d}{dt}=\frac{1}{2}\left(\frac{d_+}{dt}\bigg |_{\delta t}+\frac{d_-}{dt}\bigg |_{\delta t}\right)-\frac{i}{2}\left(\frac{d_+}{dt}\bigg |_{\delta t}-\frac{d_-}{dt}\bigg |_{\delta t}\right), 
\end{eqnarray*}
which can be used to define the {\it complex velocity}:
\begin{eqnarray}
\label{complexv}
\mathcal V=\frac{\hat d}{dt}{\bf r}=\frac{{\bf v}_++{\bf v}_-}{2}-i\frac{{\bf v}_+-{\bf v}_-}{2}={\bf V}-i{\bf U},
\end{eqnarray}
where $\bf V$ can be regarded as the classical velocity and $\bf U$ is an additional term, the {\it kink velocity} \cite{MHTeh2017}.

If we now consider the action of $\frac{\hat d}{dt}$ on a classical field $h({\bf r},t)$ under the restriction that the residual between a path and its description at a set resolution-scale is a Wiener process characterized by a diffusion constant $\mathcal D$, then we find \cite{nottale2011,MHTeh2017}:
\begin{eqnarray}\label{covardiff}
\frac{\hat d}{dt}h=\left[\frac{\partial }{\partial t}+\mathcal V\cdot \nabla -i \mathcal D\Delta \right]h.
\end{eqnarray}

The real part is the usual material derivative and the imaginary part, proportional to the Laplacian of the field, appears as a consequence of the non-differentiability. This additional term affects the Leibniz product rule, which, for two functions $f$ and $g$, becomes: 
\begin{eqnarray}\label{productrule}
\frac{\hat d}{dt}\left(f\cdot g\right)=\frac{\hat d f}{dt}g+f\frac{\hat d g}{dt}-2i\mathcal D\nabla f\cdot \nabla g.
\end{eqnarray}

If we now assume that mechanical systems involving non-differentiable or scale dependent paths can be characterized by a complex Lagrange function $\mathcal L({\bf r},\mathcal V,t)$, the enforcement of the stationary action principle with the complex action $\mathcal S=\int_{t_1}^{t_2}\mathcal L({\bf r},\mathcal V,t)dt$ leads to the usual Euler-Lagrange equation but with the complex time-differential operator and velocity. 
Considering a point  particle of mass $m$ under the influence of a real potential energy term $\Phi$, we may write the Lagrange function using its classical form with the complex velocity in the place of the usual velocity:  $\mathcal L=\frac{1}{2}m\mathcal V^2-\Phi$. Then, the Euler-Lagrange equation leads to  a generalized form of Newton's relation of dynamics where the velocity is replaced by the {\it complex velocity} (Equation \ref{complexv}) and the time derivative is replaced by the {\it complex finite time differential operator}  (Equation \ref{covardiff}):
\begin{eqnarray*}
m\frac{\hat d\mathcal V}{dt}=-\nabla\Phi.
\end{eqnarray*}

Using Equation \ref{covardiff}, this relation of dynamics can be expanded to
\begin{eqnarray}\label{gennewton}
m\frac{\partial\mathcal V}{\partial t}+m (\mathcal V\cdot \nabla) \mathcal V-im\mathcal D\Delta\mathcal V=-\nabla\Phi,
\end{eqnarray}
and, defining the complex momentum $\mathcal P=m\mathcal V=\nabla \mathcal S$, and using the identities, $(\mathcal V\cdot \nabla)\mathcal V= \frac 1 2 \nabla \mathcal V^2-\mathcal V\times(\nabla\times\mathcal V)$ and $\Delta\mathcal V=\nabla(\nabla\cdot\mathcal V)-\nabla\times(\nabla\times\mathcal V)$, while noting that, since $\mathcal V$ is a gradient, $\nabla\times\mathcal V=0$, we obtain that all the terms appear as gradients and the equation can be integrated to

\begin{eqnarray}\label{hamiltonjacobi}
\frac{\partial \mathcal S}{\partial t}+\frac{1}{2}m\mathcal V^2-im\mathcal D \nabla\mathcal V+\Phi=0, 
\end{eqnarray}
where the integration constant is absorbed in the potential energy $\Phi$. We recognize this equation as a complex Hamilton-Jacobi equation, in which we identify the complex kinetic energy as
\begin{eqnarray}\label{kineticnrg}
\mathcal T=\frac{1}{2}m\mathcal V^2-im\mathcal D \nabla\mathcal V. 
\end{eqnarray}

Then, defining a function $ \psi$ by writing the action $\mathcal S=-i \mathcal S_0 \ln \psi$, the Hamilton-Jacobi Equation \ref{hamiltonjacobi} takes the form of a Schr\"odinger equation for $\psi$ with $\hbar$ replaced by $\mathcal S_0$ for the natural choice $\mathcal S_0=2m\mathcal D$ \cite{nottale2011,MHTeh2017}:
\begin{eqnarray*}\label{schrodingerone}
 2im\mathcal D{{\partial\psi}\over{\partial t}} =-2m\mathcal D^2 \Delta\psi+\Phi\psi.
\end{eqnarray*}

The other axioms of the construction of standard quantum mechanics with, in particular, observables, Born's rule, and von Neumann's projection postulate can all be obtained or interpreted in this context as discussed elsewhere \cite{nottale2011, nottale2007, MHTeh2017}. 

It will be useful to have expressed the complex velocity in terms of the wave function written as $\psi=\sqrt{\rho}e^{i\alpha}$, which gives $\mathcal V=-2i\mathcal D\nabla\ln\psi=2\mathcal D\nabla\alpha-i\mathcal D\nabla\ln\rho$, from which we identify expressions for the classical velocity ${\bf V}=2\mathcal D\nabla\alpha$ and for the kink velocity ${\bf U}=\mathcal D\nabla\ln\rho$ \cite{saeed2020}.  Also, with the same expression for the wave function $\psi$, separating the real and imaginary parts of this Schr\"odinger equation leads to the system of Madelung equations \cite{madelung1927}: 
\begin{eqnarray}\label{schrodinger}
\frac{\partial\rho}{\partial t}&=&-\nabla\left(\rho{\bf V}\right),\label{madelungcontinuity}\\
\left(\frac{\partial}{\partial t}+{\bf V}\nabla\right){\bf V}&=&-\frac{\nabla\Phi+\nabla\mathcal Q}{m}.
\end{eqnarray}
These equations distinguish themselves from Euler's equations for inviscid fluid dynamics by the term $\mathcal Q=-2m\mathcal D^2\frac{\nabla^2\sqrt{\rho({\bf r})}}{\sqrt{\rho({\bf r})}}$ known as the quantum potential \cite{bohm1952}. The quantum potential term is responsible for the quantum nature of the solutions. The quantum potential can be expressed solely in terms of the kink velocity $\bf U$ which characterizes the non-differentiable nature of the dynamical paths, and, as such, it appears as being of a kinetic nature \cite{holland2015}: 
\begin{eqnarray}\label{qpotkinkv}
\mathcal Q_k=-\frac{1}{2}m_k{\bf U}_k^2-m_k\mathcal D\nabla\cdot{\bf U}_k
\end{eqnarray}

 We will use this relation between the quantum potential and the kink velocity while discussing the resolution-scale relativistic virial theorem in the next section as we now have all we need for this.

\section{Resolution-scale relativistic virial}\label{srvirialsec}
Using the complex momentum and velocity of each particle $\mathcal P_k=m_k\mathcal V_k$, we  can define the scale-relativistic virial as: 
\begin{eqnarray*}
\mathcal G=\sum\limits_{k=1}^N {\bf r}_k{\mathcal P}_k=\sum\limits_{k=1}^N m_k{\bf r}_k{\mathcal V}_k.
\end{eqnarray*}
Applying the complex time differential defined by Equation \ref{covardiff} following the product rule of Equation \ref{productrule}, while using the identifications $\frac{\hat d {\bf r}_k}{dt}=\mathcal V_k$ and $m_k\frac{\hat d\mathcal V_k}{dt}={\bf F_k}$ from Equation \ref{gennewton}, we obtain:
\begin{eqnarray*}
\frac{\hat d\mathcal G}{dt}=\sum\limits_{k=1}^N \left(m_k\mathcal V_k^2-2m_k i\mathcal D\nabla\cdot\mathcal V_k+{\bf r}_k\cdot{\bf F}_k\right).
\end{eqnarray*}
In this expression, we identify (see Equation \ref{kineticnrg}) the sum of twice the total complex kinetic energy $\mathcal T_{Tot}=\sum\limits_{k=1}^N\mathcal T_k$, in such a way the expression of the resolution-scale relativistic virial energy $\frac{\hat d\mathcal G}{dt}$ has the same form as its classical counter part $\frac{dG}{dt}$. With $\overline{\frac{\hat d\mathcal G}{dt}}=0$, we obtain the resolution-scale relativistic virial theorem:
\begin{eqnarray*}
2\overline{\left(\mathcal T_{Tot}\right)}=-\overline{\left(\sum\limits_{k=1}^N {\bf r}_k\cdot{\bf F}_k\right)}.
\end{eqnarray*}

It has the same form as the classical virial theorem of Equation \ref{classvirial}. It is however desirable to express the left-hand side in a more familiar way, which will hopefully reveal the classical and quantum components.  For this we may notice that the right-hand side is real and so we focus on the real part of the complex kinetic energy in the form of Equation \ref{kineticnrg} using Equation \ref{complexv}: 
\begin{eqnarray*}
\Re\left[\mathcal T_{Tot}\right]&=&\sum\limits_{k=1}^N m_k\left({\bf V}_k^2-{\bf U}_k^2-2 \mathcal D\nabla\cdot {\bf U}_k\right).
\end{eqnarray*}

We recognize the usual kinetic energy and the quantum potential in terms of the kink-velocity (Equation \ref{qpotkinkv}) and the resolution-scale relativistic virial theorem takes the form
\begin{eqnarray}\label{srvirial}
2\overline{\left(\mathcal T_{Tot}\right)}=2\overline{\left(T_{Tot}\right)}+2\overline{\left(\mathcal Q_{Tot}\right)}=-\overline{\left(\sum\limits_{k=1}^N{\bf r}_k\cdot{\bf F}_k\right)}, 
\end{eqnarray}
where we have introduced $\mathcal Q_{Tot}$, the total quantum-like potential. 

The classical virial theorem of Equation \ref{classvirial} is recovered when ${\bf U}_k=0$ and therefore $\mathcal Q_{Tot}=0$. 

Inversely, in the case $\bf V_k=0$, as in an orbital quantum ground state, we have $T_{Tot}=0$ and all the kinetic energy is in the form of the quantum potential $\overline{\left(\mathcal T_{Tot}\right)}=\overline{\left(\mathcal Q_{Tot}\right)}$. It is however important to note that it is only in this specific case of {\it motion-less dynamics} that $\mathcal T_{Tot}=\mathcal Q_{Tot}$. Indeed, more generally, starting from the expression for the complex kinetic energy in Equation \ref{kineticnrg} and proceeding with the substitution $\mathcal V_k=-2i\mathcal D\nabla\ln\psi_k$, we obtain the usual expression for the quantum kinetic energy provided the identification $\hbar\leftrightarrow 2m_k\mathcal D$,
\begin{eqnarray}\label{quantkinen}
\mathcal T_k{\psi_k}=-\frac{\left(2m_k \mathcal D\right)^2}{2m_k}{\nabla^2\psi_k}, 
\end{eqnarray}
and the resolution-scale relativistic kinetic energy $\mathcal T_{Tot}$ is the same as the standard quantum mechanical $\hat T_{Tot}$
appearing in the quantum virial theorem of Equation \ref{quantvirialeq}. In fact, we wrote Equation \ref{quantkinen} under a form that makes it ressemble an eigenvalue equation: the action of a hermitian differential operator $\hat T_k=-\frac{\left(2m_k \mathcal D\right)^2}{m_k}{\Delta}$ onto $\psi_k$ generally amounts to multiplying $\psi_k$ by a complex function of coordinates $\mathcal T_k$, which is real and constant only for those eigen-wave-functions $\psi_k$ corresponding to eigen-kinetic-energies  $\mathcal T_k$. The resolution-scale relativistic virial theorem expressed by Equation \ref{srvirial} regroups both the classical and the standard quantum mechanical cases under a single time averaging $\overline{\left(\cdots\right)}$.

\section{Summary and conclusion}\label{sumconc}

We have established the virial theorem in the resolution-scale relativity framework for non-differentiable paths described by a Wiener process characterized by a diffusion constant $\mathcal D$. Just as in classical mechanics, the resolution-scale relativistic virial theorem takes the form of an equation between the time average of the same function of the interactions between the system constituents  and the time average of the kinetic energy. As such, when restricted to differentiable paths, it reproduces the classical mechanics form of the virial theorem. 

Under the identification $\hbar\leftrightarrow 2m\mathcal D$, the resolution-scale relativistic virial theorem also reproduces the quantum mechanical form of the virial theorem which extends the classical form by applying the time averaging to quantum expectation values. This ability of the resolution-scale relativity virial theorem  to also apply to the quantum regime, while only involving a time averaging, follows from the kinetic energy to include an additional term recognized as a quantum potential. We have seen that this quantum potential appears as a misnomer as it fundamentally is of a kinetic nature since it can be expressed in terms of the kink-velocity associated with the non-differentiable component of the dynamical paths. 

In natural sciences, we attach a great importance to the uncertainty or finite resolution with which quantitative observational data are obtained. Frequently, uncertainties are propagated through the application of theories and uncertainties are used to assess the compatibility between observational data and theoretical prescriptions. However, the fact that nature is only observed at finite resolution does not generally enter fundamental theories at the foundational level. We can note for example that the concepts of instantaneous velocity or acceleration at the center of Newtonian mechanics are entirely abstract and even experimentally out of reach as only mean velocities and accelerations can be measured in practice. In fact the concept of instantaneous velocity rests on the use of calculus differentiation, which involves taking the limit to infinitesimal time differences, which, if it were applied experimentally, would take the experiment to the quantum regime in which newtonian mechanics is inapplicable and the concept of instantaneous {\it real} velocity looses its meaning. 

Postulating the resolution-scale relativity principle is an attempt at taking into account the different finite resolutions with which we observe the universe. Enforcing this postulate in the construction of mechanics leads to incorporating non-differentiable dynamical paths and, under the restriction of the non-differentiable component to be a Wiener process, the exercise leads to relations of both Newtonian mechanics at coarser resolution scales and quantum mechanics with Schr\"odinger's equation at finer resolution scales.

Inversely, pushing classical mechanics sufficiently far to coarser resolution scales leads the realm of deterministic chaos with a loss of actual predictability. Only bundles of an infinite multitude of  effectively indistinguishable and non-differentiable dynamical paths can provide an accurate description of the system's evolution while remaining in the chaotic regime. The resolutions-scale relativity principle may then be at play in the dynamics. The non-differentiable component of the paths may not exactly be a Wiener process, which would correspond to the dynamics to be strictly memory less \cite{nottale2011, MHTeh2017} but it cannot be too far from a Wiener process as otherwise predictability would start being recovered. This suggests the possible emergence of a quantum-like dynamics with $\hbar$ replaced with some value of $2m\mathcal D$ specific to the system. If this is the case, the resolution-scale relativistic virial theorem obtained above (Equation \ref{srvirial}) would become applicable.

Astronomers measure the velocity dispersion of stars in galaxies or of galaxies in galaxy clusters. Combining these measurements with  estimates of the masses of the moving objects, they obtain estimates of the kinetic energy $T_{Tot}$. The virial theorem of Equation \ref{classvirial} is then used to obtain an estimate of the potential energy of the stars or galaxies. Indeed, the application of the classical virial theorem to a gravitationally bound system of visible mass $M_0$ and radial extent $R_0$ indicates the internal speeds to be on a scale $v_G\sim\sqrt{\frac{G M_0}{R_0}}$. Smaller speeds are expected at greater distances from the center of the system, eventually tailing off in proportion to the square root of the inverse of the distance. Observations however indicate that the speeds do not decrease as expected and maintain higher values further out. 

When the observed classical kinetic energy is assumed to be matched exclusively by  gravitational potential energy in the classical virial theorem, one is led to concluding that the actual mass and radial extent of the system both exceed the visible mass and extent. This suggests the presence of an invisible material, the Dark Matter \cite{zwicky1937,rubin1980}, responsible for the additional gravitational potential \cite{pardo2020} while the nature of the Dark Matter remains to be identified.

If the resolution-scale relativity virial theorem is applied instead, the measured average classical kinetic energy would have to be matched to the combination of the gravitational potential and the quantum-like potential:
\begin{eqnarray}
\overline{\left(T_{Tot}\right)}=-\frac{1}{2}\overline{\left(\sum\limits_{k=1}^N{\bf r}_k\cdot{\bf F}_k\right)}-\overline{\left(\mathcal Q_{Tot}\right)}. 
\end{eqnarray}
From its expression in terms of the density,  we can see that the quantum potential $\mathcal Q=-2m\mathcal D^2\frac{\nabla^2\sqrt{\rho({\bf r})}}{\sqrt{\rho({\bf r})}}$ is positive toward the center of the system where the density is maximal and negative in the outskirt of the system where the density tails off. This would amount to a reduction of the velocity scale in the central regions and to an increase in the outer regions by an amount of the order of $v_Q\sim\frac{\mathcal D}{R_0}$ where the effective diffusion coefficient $\mathcal D$ has to be established. For an order of magnitude, we can consider $v_Q\sim v_G$, from which $\mathcal D\sim \sqrt{GM_0R_0}$. With this we can estimate the de\,Broglie-like wavelength $\lambda_{dB}=\frac{2m\mathcal D}{mv_G}\sim 2R_0$. The de\,Broglie-like wavelength sets the resolution scale at which the transition between the classical and quantum-like regimes is observed. It is then not surprising to obtain that the de\,Broglie-like wavelength matches the confinement scale. We can also note here that the de\,Broglie-like wavelength does not depend on the particle's mass and objects of very different masses can share the same quantum-like wave function and confinement scale in the system they are part of. 

In order to tbe more specific regarding the virial velocity profile contribution resulting from a quantum like potential, we can consider a spherically symmetric system with a gaussian radial density profile. The quantum-like potential is a quadratic function of the distance. The corresponding virial velocity contribution is then an increasing linear function of the distance from the center. A constant quantum potential and contribution to the virial velocity would result from a radial density profile following $\rho(r)\propto\frac{e^{-2r/R_0}}{r^2}$. This would correspondingly change the conclusions reached under the Dark Matter hypothesis as a smaller amount of invisible material is then required to account for the observed speed distribution. Actual tests of the possibility that a quantum-like potential contributes to the cosmological dark potential  would require dedicated analysis of the visible mass and virial speed profiles in specific systems.   

\section*{Acknowledgments}{The author acknowledges comments and suggestions from Janvida Rou, Eugene Mishchenko, and John Belz.}

\section*{Appendix A: Classical virial theorem from quantum mechanics}\label{classicquantvirial}
Starting from the definition of $G_C$ :
 $$
 {d\over{dt}}G_C=\sum\limits_{j=1}^N\left(\left({d\over{dt}}\langle {\bf p}_j\rangle\right) \langle {\bf r}_j\rangle+\langle {\bf p}_j\rangle\left({d\over{dt}} \langle {\bf r}_j\rangle\right)\right)
 $$
 We can then apply the Ehrenfest theorem:
$$
{d\over{dt}} G_C ={1\over{i\hbar}}\sum\limits_{j=1}^N\left( \langle [{\bf p}_j,H] \rangle \langle {\bf r}_j \rangle + \langle {\bf p}_j \rangle \langle [{\bf r}_j,H] \rangle\right)
$$
And, with the Hamiltonian $H=-\sum\limits_{j=1}^N\frac{{\bf p}_j^2}{2m_j}+V_{Tot}$, we have $\frac{1}{i\hbar}[{\bf p}_j,H]=-\nabla_j V_{Tot}$ and $\frac{1}{i\hbar}[{\bf r}_j,H]=\frac{{\bf p}_j}{m_j}$, with which we obtain the virial theorem as in Equation \ref{expectvirial} by identifying ${\bf F}_j=-\nabla_j V_{Tot}$.
\section*{Appendix B: Quantum virial }\label{quantvirial}
With the canonical commutation of ${\bf r}_j$ and ${\bf p}_j$, $G_Q=\sum\limits_{j=1}^N {\bf p}_j\cdot {\bf r}_j-\frac{3N}{2}i\hbar$ and so, Ehrenfest theorem gives: 
$$
{{d \langle G_Q\rangle}\over{dt}}= {1\over{i\hbar}}\sum\limits_{j=1}^N \langle [{\bf p}_j\cdot {\bf r}_j,H] \rangle
 $$
With the Hamiltonian $H=\sum\limits_{k=1}^N \frac{{\bf p}_k^2}{2m_k}+V_{Tot}$, we can use $\frac{1}{i\hbar}[G_Q,{\bf p}_k^2]=2{\bf p}_k^2$, and $\frac{1}{i\hbar}[G_Q,V_{Tot}]= -\sum\limits_{k=1}^N{\bf r}_k\cdot\nabla_k V_{Tot}$
 and so, with $T_{Tot}=\sum\limits_{k=1}^N\frac{{\bf p}_k^2}{2m_k}$, and ${\bf F}_k=-\nabla_k V_{Tot}$, we have: 
 $${{d \langle G_Q\rangle}\over{dt}}=2\langle T_{Tot}\rangle+\langle\sum\limits_{k=1}^N {\bf r}_k\cdot {\bf F}_k\rangle,$$ which leads to the quantum virial theorem expressed by Equation \ref{quantvirialeq}.

\end{document}